 \documentclass[12pt]{article}
\usepackage{amsmath}

\makeatletter \@addtoreset{equation}{section} \makeatother
\renewcommand{\theequation}{\thesection.\arabic{equation}}
\newcounter{alfa}
\newcounter{alfax}
\def\text#1{\mbox{#1}}
\def\equ#1{(\ref{#1})}

\def\bea{\setcounter{alfa}{0}\setcounter{alfax}{1}\addtocounter{alfax}{\value{equation}}
\renewcommand{\theequation}{\addtocounter{alfa}{1}\thesection.\arabic{alfax}\alph{alfa}}
\begin{eqnarray}}
\def\eea{\end{eqnarray}\setcounter{equation}{\value{alfax}}
\setcounter{alfax}{0}\renewcommand{\theequation}{\thesection.\arabic{equation}}}
\def\be#1{\begin{equation}\label{#1}}
\def\ee{\end{equation}}
\def\equ#1{(\ref{#1})}

\def\cal{\mathcal}

\begin{document}

\title{Duality and fields redefinition in three dimensions}
\author{M\' arcio A. M. Gomes\thanks{Electronic address: gomes@fisica.ufc.br} and R. R. Landim\thanks{Electronic address: renan@fisica.ufc.br}\\
Universidade Federal do Cear\'{a} - Departamento de F\'{\i}sica \\
C.P. 6030, 60470-455 Fortaleza-Ce, Brazil}
\maketitle

\begin{abstract}
We analyze local fields redefinition and duality for gauge field theories in three dimensions.
We find  that both Maxwell-Chern-Simons and the Self-Dual models admits the same fields
redefinition. Maxwell-Proca action and its dual also share this property. We show explicitly that a gauge-fixing term has no influence on duality and fields redefinition.
\end{abstract}
\vspace{1.0cm}


\vspace{0.3cm}


\section{Introduction}

In this paper we investigate the relationship between duality and
fields redefinition, in a sense introduced in \cite{sorella1} and
\cite{sorella2}, for Maxwell-Chern-Simons and Maxwell-Proca models in three space-time dimensions.

It is widely known that a sufficient condition for duality is the
existence of a global symmetry in the original action and one
finds the dual model "gauging" this symmetry \cite{quevedo}. The
original model is now seen as a gauge-fixed version of its dual
model. One secure prescription for achieving that is "gauge
embedding" procedure \cite{clovis1}.

This formalism can be used to derive the well-known duality
between self-dual model (SD) and Maxwell-Chern-Simons action (MCS)
\cite{Deser1}. Some questions are natural consequences of this
fact. The first one: this duality implies that both models will
have the same field redefinition? The answer is yes, and we prove.
A second question is if this property holds to some other dual
gauge models. To provide an example, we deal with Maxwell-Proca
action (MP) and its dual to show that they indeed have the same
field redefinition. A more rigorous treatment about this apparent
connection between duality and fields redefinition is given in
\cite{marcio2}.

This paper is organized as follows. In Section 2 we prove that SD
and MCS models have the same fields redefinition.  Section 3 is
devoted to establish the same fact MP and its dual. In appendix A
we detail the rules for functional calculus with differential
forms.

We use the following conventions: the metric adopted is $(-++)$.
We omit the wedge product symbol $\wedge $ and use the inner
product of two $p$-forms $\left( \omega _{p},\eta _{p}\right)
=\int \omega _{p}\wedge ^{\ast }\eta _{p}$. $\ast $ denotes the
usual duality Hodge operator.

\section{Field Redefinitions on Maxwell-Chern-Simons (MCS) and Self-Dual (SD) Models \qquad}

Twenty years ago, Deser and Jackiw \cite{Deser1} established
duality between (SD) action
\begin{equation}
S_{SD}=\int \left( \frac{m^{2}}{2}A^{\ast }A+\frac{1}{2}mAdA\right) =\frac{%
m^{2}}{2}\left( A,A\right) -\frac{m}{2}\left( A,^{\ast }dA\right),\label{SD}
\end{equation}
and topologically massive MCS action
\begin{equation}
S_{MCS}=\int \frac{1}{2}dA^{\ast }dA-\frac{m}{2}AdA=\frac{1}{2}\left(
dA,dA\right) +\frac{m}{2}\left( A,^{\ast }dA\right),\label{MCS}
\end{equation}
following the clue that both models involve a massive vectorial
field in three dimensions and violate parity. They derive the
duality from a master action. Each model is obtained combining
equations of motion and this master action. This can be also achieved using gauge embedding formalism
\cite{clovis1}.

In \cite{sorella1} and \cite{sorella2} Lemes \textit{et al} showed that
Chern-Simons term can be seen as a generator for MCS model, through a field
redefinition. We set the redefinition for
\begin{equation}
\frac{1}{2}\left( dA,dA\right) +\frac{m}{2}\left( A,\ast dA\right) =\frac{%
m}{2}\left( \hat{A},\ast d\hat{A}\right),  \label{A02}
\end{equation}
where
\begin{equation}
\hat{A}=A+\sum_{i=1}^{\infty }\frac{A_{i}}{m^{i}}.  \label{ARED}
\end{equation}
It was shown in \cite{marcio1} that
\bea
\delta A_{i} &=&0, \\
d^{\dagger }A_{i} &=&0,
\eea
and that $A_{i}$ depends linearly on $A$ in the following way
\begin{equation}
A_{i}=\alpha _{i}\left( \ast d\right) ^{i}A. \label{A04}
\end{equation}
 We can also look at field redefinition \equ{ARED} as
having an operatorial nature\footnote{Let us comment about a functional formulation with this field redefinition. Since this redefinition is linear, the functional measure gains a factor which does not depend on the field and hence is factorable from the functional.},
\begin{equation}
\hat{A}=\mathcal{O}A,\label{hatA}
\end{equation}
where $\mathcal{O}$ is an operator defined by \be{O}
\mathcal{O}=\left( 1+\sum_{j=1}^{\infty }\frac{\alpha
_{j}\left( \ast d\right) ^{j}}{m^{j}}\right). \ee The operator
$\mathcal{O}$ has the following properties\bea
(\cal{O}A,B)&=&(A,\cal{O}B),\label{prop-O}\\
\left[~\cal{O},\ast d~\right]&=&0,
\eea
for any one-forms $A$ and $B$. So, we can write the field redefinition for the MCS model in
the form
 \be{MCS1}
\frac{1}{2}\left( dA,dA\right) +\frac{m}{2}\left( A,\ast dA\right) =\frac{%
m}{2}\left( \cal{O}{A},\ast d\cal{O}{A}\right).
 \ee
Taking the functional derivative with respect to $A$ on both sides of this equation we obtain
\begin{equation}
\left( \frac{\ast d}{m}-1+\mathcal{O}^2\right) \ast dA=0.
\label{ODA}
\end{equation}
Since the operator $\cal{O}$ depends only on $\ast d$, we must
have
\begin{equation}
\cal{O}^2=1-\frac{\ast d}{m}\label{Ofun},
\end{equation}
and consequently
\begin{equation}
\cal{O}^2A=A-\frac{\ast d}{m}A\label{O2A}.
\end{equation}
Integrating with respect to $A$ we get
\begin{equation}
\frac{1}{2}m^{2}(A,A)-\frac{1}{2}m\left( A,\ast dA\right) =\frac{1}{2}%
m^{2}\left( \hat{A},\hat{A}\right).\label{SD1}
\end{equation}
So the fields redefinitions to MCS and its dual SD model are the same
 and SD model can be reset in a pure mass-like term. From \equ{Ofun} we can obtain the form
of the operator $\cal{O}$ by expansion in power series in $\ast
d/m$
\begin{equation}
\mathcal{O}=1-\sum_{j=1}^{\infty }\frac{\left( 2j\right) !}{%
2^{2j}\left( 2j-1\right) \left( j!\right) ^{2}}\left( \frac{\ast d}{m}%
\right) ^{j}.  \label{Oexp}
\end{equation}
Comparing \equ{hatA} and \equ{Oexp}, we find
out a  formula for the coefficients $\alpha _{j}\prime s$
\begin{equation}
\alpha _{j}=-\frac{\left( 2j\right) !}{2^{2j}\left( 2j-1\right) \left(
j!\right) ^{2}}\label{alphas},
\end{equation}
that furnishes the same coefficients found in \cite{sorella4}.

We can directly invert the field redefinition
\begin{equation}
A=\mathcal{O}^{-1}\hat{A},
\end{equation}
with
\begin{equation}
\mathcal{O}^{-1}=\left( 1-\frac{\ast d}{m}\right)
^{-1/2}=\sum_{k=0}^{\infty }\frac{\left( 2k\right) !}{2^{2k}\left( k!\right)
^{2}}\left( \frac{\ast d}{m}\right) ^{k},
\end{equation}
that are also in complete agreement with that ones \cite{sorella4}.

The series expansion \begin{equation}
(1+x)^{1/2}=\sum_{j=1}^{\infty }(-1)^{j}\alpha_{j}x^{j},
\end{equation}  converges only for $\vert{x}\vert<1$ and $x=1$, but this is not a problem since one can easily extend the convergence for another intervals by analytic continuation.

We can change our way of thinking and consider  as \equ{ODA} had
been obtained from \equ{O2A} after application of $\ast d$. Thus
redefined MCS is generated by redefined SD model, both presenting
the same field redefinition. One can reasonably argue what would
one get applying $\ast d$ twice on \equ{O2A} and integrating on
field $A$. We obtain a new gauge invariant model
\begin{equation}
-\frac{1}{2}\left( dA,d\ast dA\right) +\frac{m}{2}\left( dA,dA\right) =\frac{%
m}{2}\left( d\hat{A},d\hat{A}\right).   \label{2dsd}
\end{equation}
By construction  the field redefinition of \equ{2dsd} is
exactly that for SD and MCS models. In fact, these models belongs to a same equivalence class
\cite{marcio2}.

 An interesting point is that adding a gauge-fixing term to SD model
 have no influence on fields redefinition since that it does not spoil duality.
Consider the Self-Dual model plus a Landau gauge fixing term
\be{sd-gf3} S^{(0)}=\int \left( \frac{m^2}{2}A_\mu
A^\mu+\frac{\alpha}{2}(\partial_\mu A^\mu)^2-
\frac{m}{2}\epsilon_{\mu\nu\rho}A^\rho\partial^\mu
A^\nu\right)d^3x. \ee Following the gauge embedding prescription
\cite{clovis1}, the first iterative action is \be{s1-3}
S^{(1)}=S^{(0)}-\int\left(m^2A_\mu-\alpha\partial_\mu(\partial
A)-m\epsilon_{\mu\nu\rho}\partial^\nu A^\rho\right)B^\mu d^3x, \ee
and it follows that \be{deltas1-3} \delta S^{(1)}=-\delta
\int\frac{m^2}{2}B_\mu B^\mu d^3x
-\delta\int\frac{\alpha}{2}(\partial_\mu B^\mu)^2d^3x. \ee Then,
the invariant action is \be{inv3}
S^{(2)}=S^{(1)}+\int\left(\frac{m^2}{2}B_\mu
B^\mu+\frac{\alpha}{2}(\partial_\mu B^\mu)^2\right)d^3x. \ee The
equation of motion for $B_\mu$ shows that $\partial_\mu
B^\mu=\partial_\mu A^\mu$ and
$m^2B_\mu=m^2A_\mu-m\epsilon_{\mu\nu\rho}\partial^\nu A^\rho$.
This points to the Maxwell-Chern-Simons as being the dual model no
matter what the gauge fixing term is.

\section{Fields Redefinitions on Maxwell-Proca (MP) and its Dual \qquad}

Another model which shares the same fields redefinition with its dual is Maxwell-Proca (MP),
\be{proca}
S_{MP}=\frac{m^2}{2}(A,A)-\frac{1}{2}(\ast dA,\ast
dA)=\frac{m^2}{2}(A,\left(1-(\ast d/m)^2\right)A).\ee
We can find its dual following the gauge embedding procedure,
\be{procadual} S_{dual-MP}=\frac{1}{2}(\ast
dA,\ast dA)-\frac{1}{2m^2}(\ast d\ast dA,\ast d\ast dA). \ee
Just like SD model, MP can be redefined to a pure masslike term namely
\be{proca-redef} S_{MP}=\frac{m^2}{2}(\hat{A},\hat{A})=\frac{m^2}{2}(OA,OA), \ee
where $O$ also satisfies the properties (\ref{prop-O}). From (\ref{proca}) and (\ref{proca-redef}), we see that $O$ is nothing but:
\begin{equation}
O=(1-(\ast d/m)^2)^{1/2}.
\end{equation}
Repeating the steps in section 2, we are left with:
\begin{equation} \label{O2eq}
[O^{2}-1+(\frac{*d}{m})^2]A=0.
\end{equation}
Applying $(*d)^2$ and integrating in $A$:
 \begin{equation}
S_{dual-MP}=\frac{1}{2}(\ast
dA,\ast dA)-\frac{1}{2m^2}(\ast d\ast dA,\ast d\ast dA)=\frac{1}{2}(\ast d\hat{A},\ast
d\hat{A}).
\end{equation}
Observe that applying $*d$ in (\ref{O2eq}) and integrating, one gets another theory with same field redefinition as MP that is not its dual. We clear this point in \cite{marcio2}.
\section{Conclusion}
In this work we analysed the aspects of field redefinition and duality of gauge theories in three dimensions.
We showed that the MCS and SD models can be rewritten as a unique term with the same field redefinition.
Indeed both models are members of a equivalence class of gauge theories. Let us remark that such redefinition
is local and has a closed expression in operator $\ast d$, e.g, see equation \equ{Ofun}. The expansion in power
of  $\ast d$ is formal and only makes sense only operating on a gauge field.
Our results clarify some aspects of duality in three dimension. In this approach, the gauge fixing term drops out the dual theory.
\vskip1cm
  \noindent
  {\large\bf Acknowledgments}
  \vskip0.5cm
Conselho Nacional de Desenvolvimento Cient\'\i fico e tecnol\'ogico-CNPq is gratefully
acknowledged for financial support.

\appendix
\section{Functional calculus with differential forms}
We start by defining the functional derivative of a $p$-form
$A(x)$. Let the inner product of two p-form be given by \be{inner}
(A,B)=\int A(x)\ast
B(x)=\int\frac{1}{p!}A(x)_{\mu_1..\mu_p}B(x)^{\mu_1..\mu_p}d^Dx,
\ee where we are considering a flat manifold. Taking the
functional derivative with respect to $A(y)_{\nu_1..\nu_p}$ on
both sides of \equ{inner}, one gets \be{Ader} \frac{\delta}{\delta
A(y)_{\nu_1..\nu_p}}(A,B)=B(y)^{\nu_1..\nu_p}. \ee Then we define
the functional derivative of a $p$-form $A(x)$ as \be{Afunc}
\frac{\delta}{\delta A(y)}(A,B)=B(y). \ee We also define the Dirac
delta $p$-form \be{deltaDiracp} (\delta^D_p(x-y),B(x))=B(y), \ee
then \be{AAfunc} \frac{\delta A(x)}{\delta A(y)}=\delta^D_p(x-y).
\ee From definition \equ{deltaDiracp}, we can get the explicit
form of $\delta^D_p(x-y)$ in terms of usual Dirac delta function:
\begin{eqnarray}
\delta^D_p(x-y)=\frac{1}{p!}\delta^D(x-y)g_{\mu_1\nu_1}..g_{\mu_p\nu_p}dx^{\mu_1}
\wedge..\wedge dx^{\mu_p}\otimes dy^{\nu_1}\wedge..\wedge dy^{\nu_p}.
\end{eqnarray}
Clearly, $\delta^D_0(x-y)=\delta^D(x-y)$. Note that a $p$-form is
defined in a point of the co-tangent space, therefore two
$p$-forms in different points always commute since they are
defined in different co-tangent spaces, i.e, $B(x)B(y)=B(y)B(x)$.

\end{document}